\documentclass[a4paper]{jpconf}

\usepackage{cite}
\usepackage{graphicx}
\usepackage{subfigure}
\usepackage{xspace}
\usepackage{graphicx}
\usepackage{amssymb}
\usepackage{wasysym}
\usepackage{color}
\usepackage{lineno}
\usepackage{pgf}
\usepackage{float}
\usepackage{rotating}
\usepackage{dcolumn}
\usepackage{wrapfig}

\usepackage{citesort}

\newcommand{\jpsi}{\ensuremath{{J/\psi}}\xspace}
\newcommand{\pT}{\ensuremath{p_{\rm T}}\xspace}
\newcommand{\dEdx}{d$E$/d$x$\xspace}

\newcommand{\mee}{$\ensuremath{m_{ee}}$\xspace} 

\begin{document}

\title{Low-mass dielectron measurement in pp and Pb--Pb collisions in ALICE}

\author{Markus K. K\"{o}hler
	for the ALICE Collaboration
	}

\address{Research Division and ExtreMe Matter Institute EMMI,\\ GSI Helmholtzzentrum f\"{u}r Schwerionenforschung, Germany}

\ead{m.koehler@gsi.de}

\begin{abstract}
We report on the first dielectron measurement in pp collisons at $\sqrt{s} = 7$~TeV with the ALICE detector system. The results are compared to the 
expected hadronic sources. The hadronic cocktail agrees to the measured dielectron continuum within statistical and systematic uncertainties. 
The status of the dielectron measurement in Pb--Pb collisions at \mbox{$\sqrt{s_{NN}}=2.76$}~TeV is addressed.
\end{abstract}

\section{Introduction}
Low-mass dielectrons are an important probe for the hot and dense medium which can be created 
in ultrarelativistic heavy-ion collisions~\cite{RappWambachHees,Linnyk_dielectron}. Since leptons do not interact strongly, 
they carry the information from all collision stages with negligible 
final-state interaction. Potential modifications of the medium, such as chiral symmetry restoration,
could have measurable impact on the emission pattern of dielectrons, 
where measurements in proton-proton (pp) collisions are important as a reference for the studies in nucleus-nucleus collisions. \\
ALICE~\cite{ALICE1} provides unique capabilities at the LHC for particle tracking and electron identification at low momenta.
With ALICE, electrons can be identified at
mid-rapidity in the central barrel ($|\eta| \lesssim 0.9 $) exploiting their specific energy loss in the Inner Tracking System (ITS)
and the Time Projection Chamber (TPC), with the Transition Radiation Detector (TRD), and due to their velocity as measured with 
the time-of-flight detector (TOF). \\ 
This writeup reports on the measurement of the production cross section of dielectrons in the mass range 
$0<$~\mee~$<3.3$~GeV/$c^2$ for pp collisions at $\sqrt{s}=7$~TeV. The invariant mass distribution 
will be compared to the expected hadronic sources.\\
Moreover, the status of the analysis of Pb--Pb collisions at $\sqrt{s_{NN}}=2.76$~TeV is presented.
\section{Data analysis}
\begin{figure}[t]
\begin{minipage}{14pc}
\includegraphics[scale = 0.41]{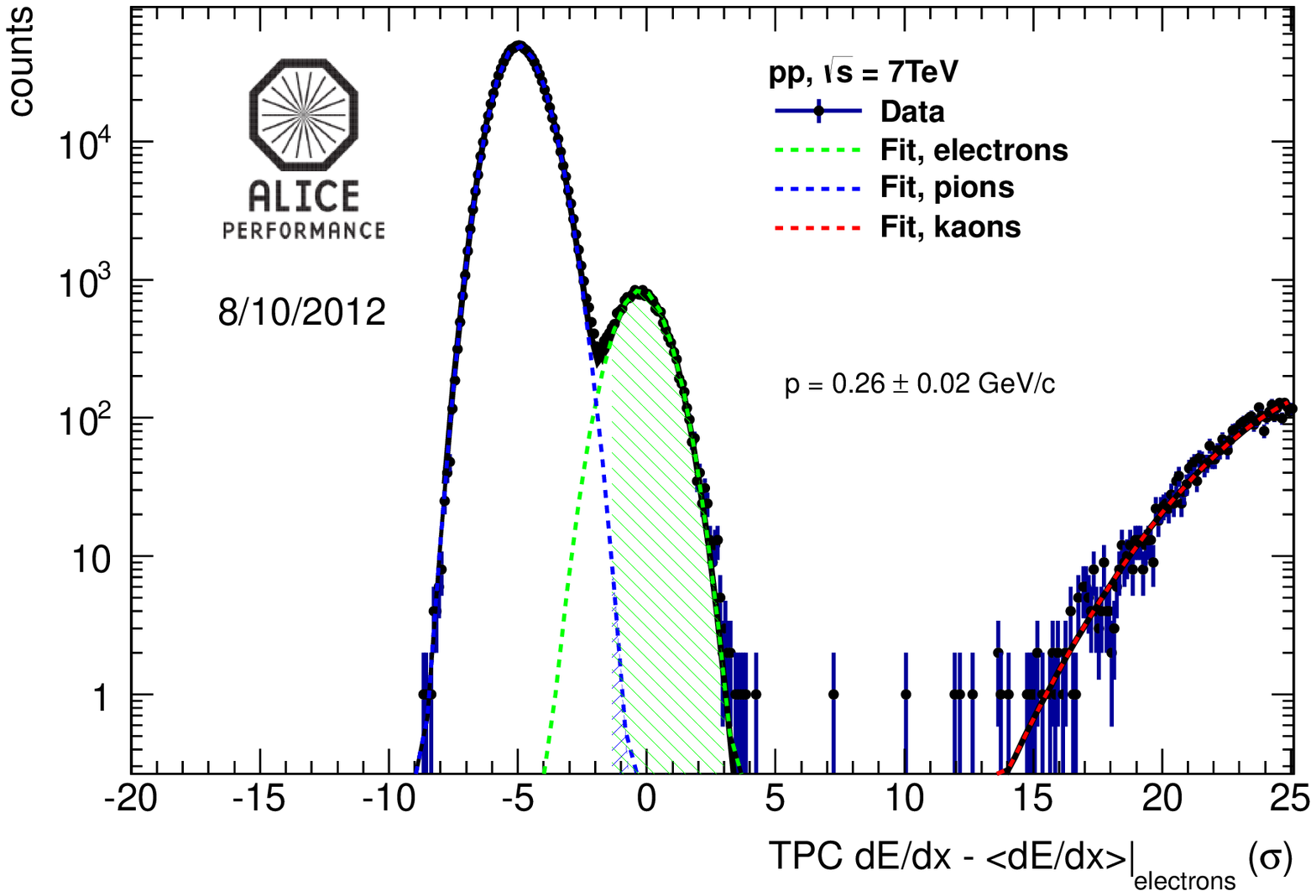}
\end{minipage}\hspace{4pc}%
\begin{minipage}{14pc}
\includegraphics[scale = 0.43]{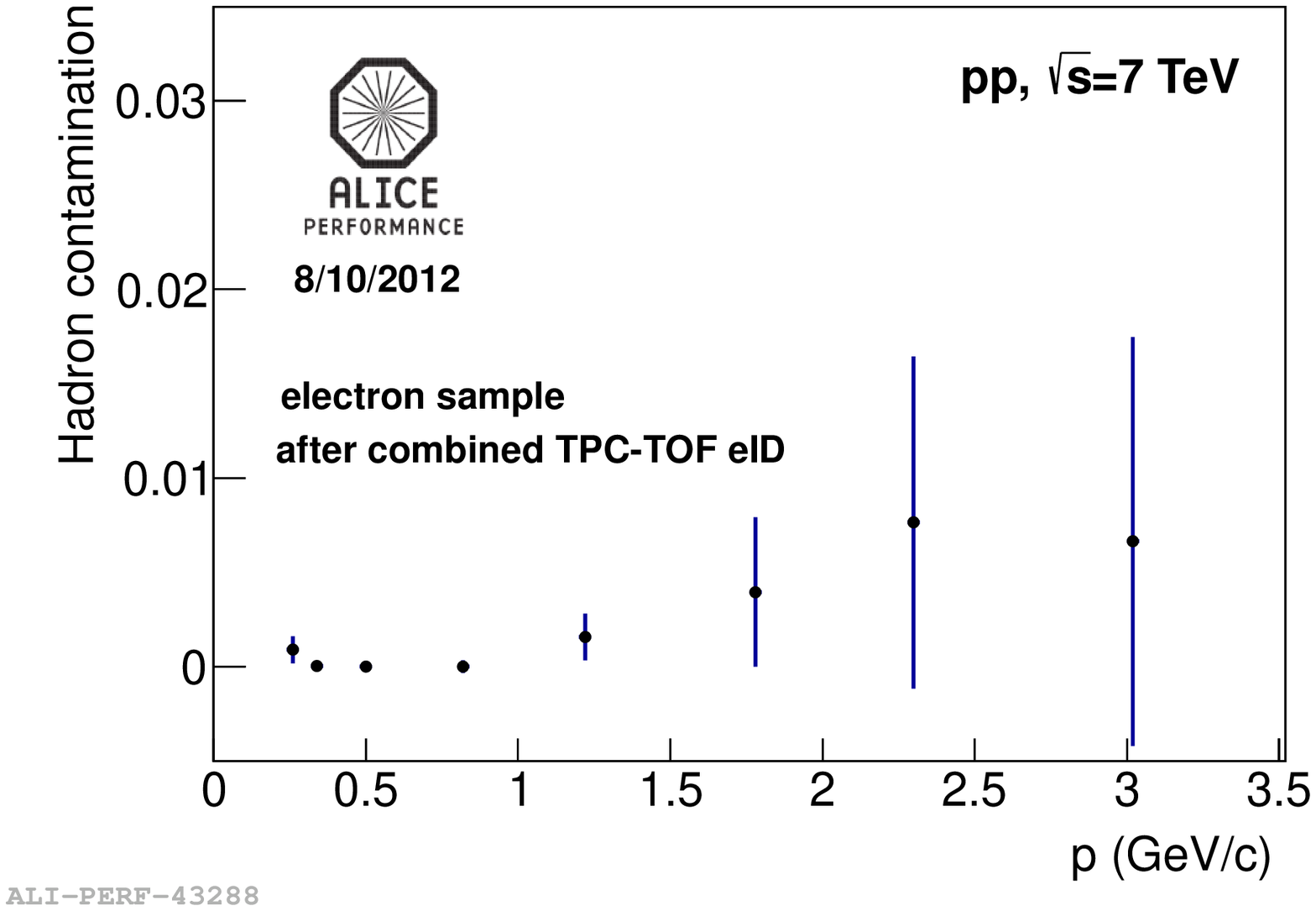}
\end{minipage} 
\caption{Estimate of the hadron contamination in the electron candidate sample. The \dEdx distributions were fitted with Gaussians. The 
integral of the fits result in a contamination of at most~1\%.}\label{fig:contamination}
\end{figure}

\begin{figure}[t]
\begin{minipage}{18pc}
\includegraphics[scale = 0.43]{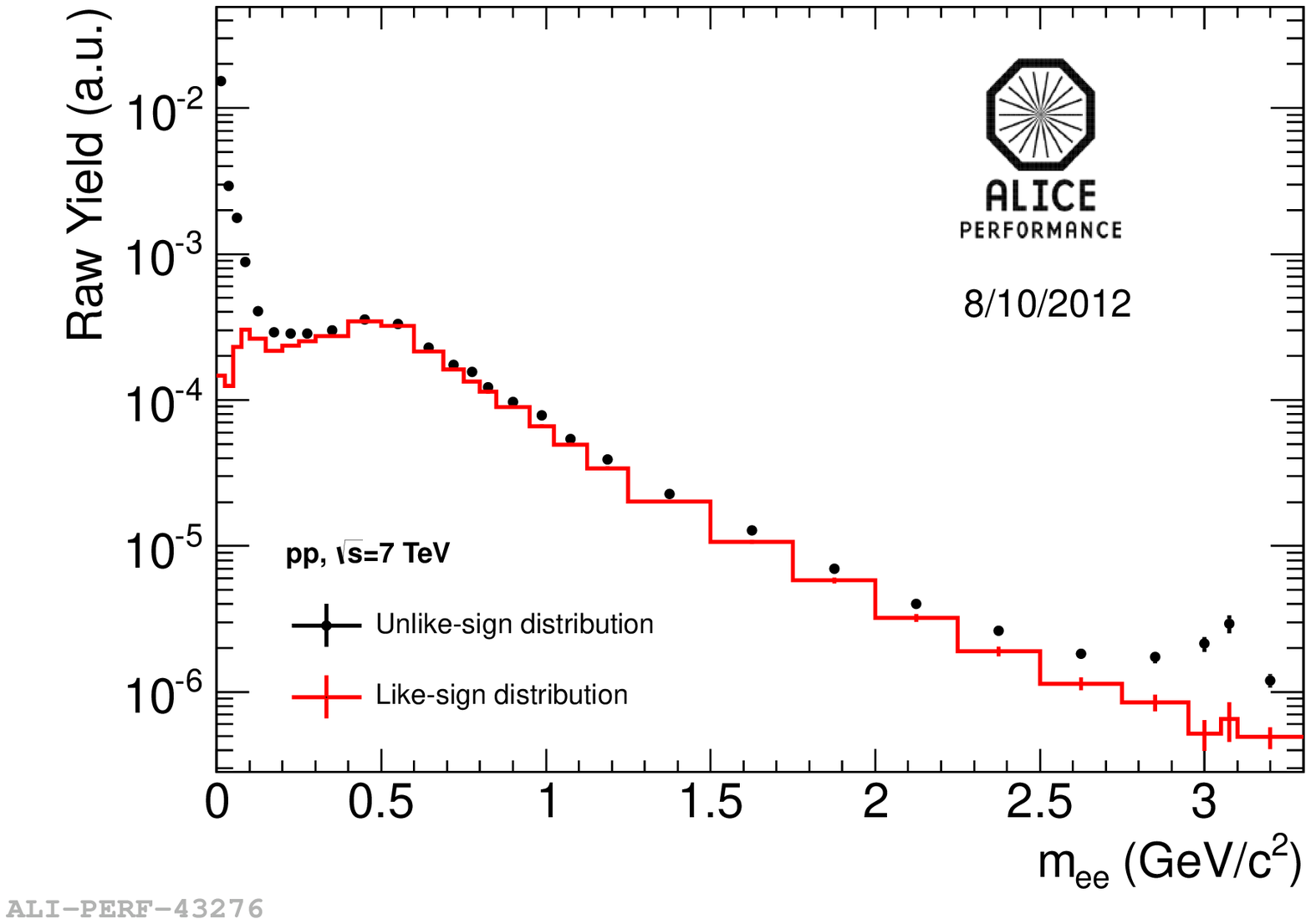}
\caption{Invariant mass distribution of unlike-sign pairs shown together with combinatorial background pairs measured via like-sign pairs 
for pp collisions. }\label{fig:unlike_like}
\end{minipage}\hspace{2pc}%
\begin{minipage}{18pc}
\includegraphics[scale = 0.4]{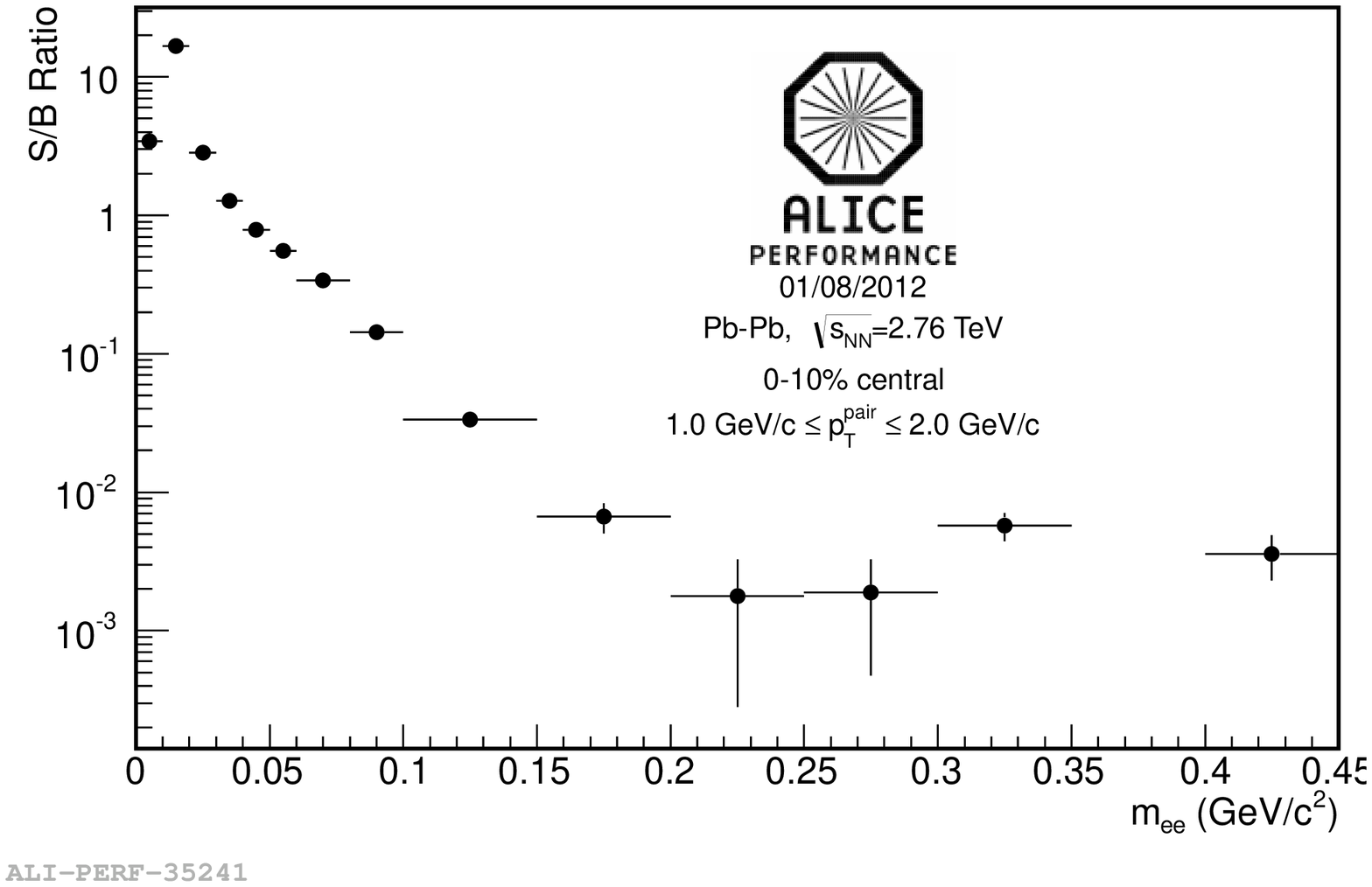}
\caption{The $S/B$ ratio for Pb--Pb collisions at $\sqrt{s_{NN}}=2.76$~TeV in the low-mass region for
$\pT^e>~0.4$~GeV/$c$.}\label{fig:LeadLead}
\end{minipage} 
\end{figure}

The results obtained in this analysis are based on data collected in 2010 for pp collisions and in 2011 for Pb--Pb collisions. 
The analysis was performed over the full azimuth in the electron pseudo-rapidity range $|\eta^e|<0.8$. The transverse 
momentum was $\pT^{e} > 0.2$~GeV/$c$ for pp collisions and $\pT^{e}> 0.4$~GeV/$c$ for Pb--Pb collisions.\\
Electrons were identified using the combined track information from TPC and TOF. In the whole transverse momentum range TPC information 
was used to reject pions. In the momentum range $0.4 < p < 5$~GeV/$c$, kaons and protons can be separated from electrons 
in the TOF and were efficiently rejected.\\
Figure~\ref{fig:contamination} shows a data driven method for the estimation of remaining hadron contamination in the electron candidate sample. 
The energy loss distributions of electron candidates in the TPC were fitted by Gaussian functions, left 
panel of Figure~\ref{fig:contamination}.
From the integral of the fits the hadron contamination was calculated as a function of the particle momentum, right panel of 
Figure~\ref{fig:contamination}. The hadron contamination in the electron sample is at most~1~\%.\\
Photon conversions were rejected via their displaced vertex. Moreover, a cut was applied on the orientation of the plane spanned by the momentum of 
the electron and positron of a pair with respect to the magnetic field direction in the central barrel.\\
The measured unlike-sign spectrum is thus an overlay of the signal and 
combinatorial background pairs. The combinatorial background $N^{CB}_{+-}$ was reconstructed with the same-event like-sign 
method. Under the assumption, that the physics signal consists only of unlike-sign pairs the combinatorial background was calculated as 
$N^{CB}_{+-} = 2R\times \sqrt{N_{++}N_{--}}$. Here $R$ is a correction factor for the acceptance difference between like-sign and 
unlike-sign pairs, obtained from mixed event distributions, and $N_{++}$ and $N_{--}$ are the yields of like-sign pairs of positive and 
negative particles, respectively. The $R$ factor was determined to be consistent with one within the statistical uncertainties.\\
The same-event like-sign method can describe the correlated background contributions from e.g. crossed pairing from double Dalitz decays at low masses 
or jet-like correlations at high masses. However, the drawback of the same-event like-sign method is the limited statistical accuracy 
of the combinatorial background compared to the mixed event method.\\
The unlike-sign distribution is shown together with the like-sign distribution for pp collisions in Figure~\ref{fig:unlike_like}. 
The raw signal was extracted by the subtraction of the like-sign distribution from the unlike-sign distribution and is corrected 
for efficiency. The efficiency correction was done on a track-by-track level in $(\pT,\eta,\phi)$ bins using Monte Carlo simulations.\\
The expected hadronic sources of dielectrons in the mass range $0<$~\mee~$<3.3$~GeV/$c^2$ 
were derived from $\pT$ differential cross sections of $\pi^0$, $\eta$, $\phi$ and  \jpsi 
measured with ALICE at mid-rapidity~\cite{ALICE_pi0,ALICE_phitokk,ALICE_jpsi}.
The shape of the mass distribution of dielectrons from correlated semi-leptonic decays of charmed 
mesons was calculated with PYTHIA~\cite{PYTHIA6}, which was scaled to the measured charm cross section~\cite{ALICE_ccbar}. Other contributions, 
which were not measured, such as the $\eta'$ and the $\rho$, were calculated by assuming $m_T$ scaling. The detector resolution was folded 
into the mass spectra.\\
\begin{figure}[t]
\begin{minipage}{18pc}
\includegraphics[scale = 0.43]{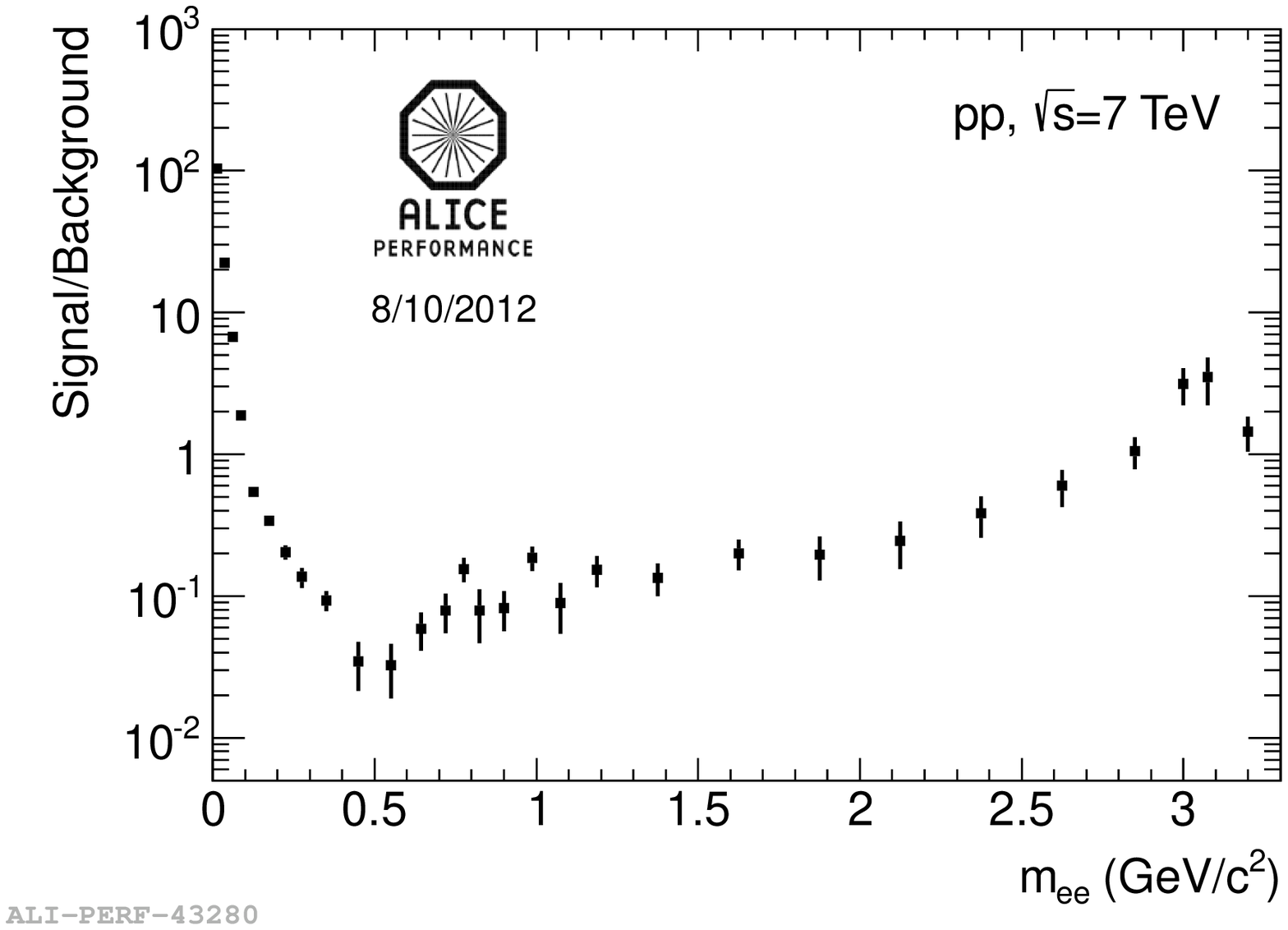}
\caption{The $S/B$ ratio for pp collisions in the mass range $0<$~\mee~$<3.3$~GeV/$c^2$.}\label{fig:pp_SoB}
\end{minipage}\hspace{2pc}%
\begin{minipage}{18pc}
\centering
 \includegraphics[scale = 0.4]{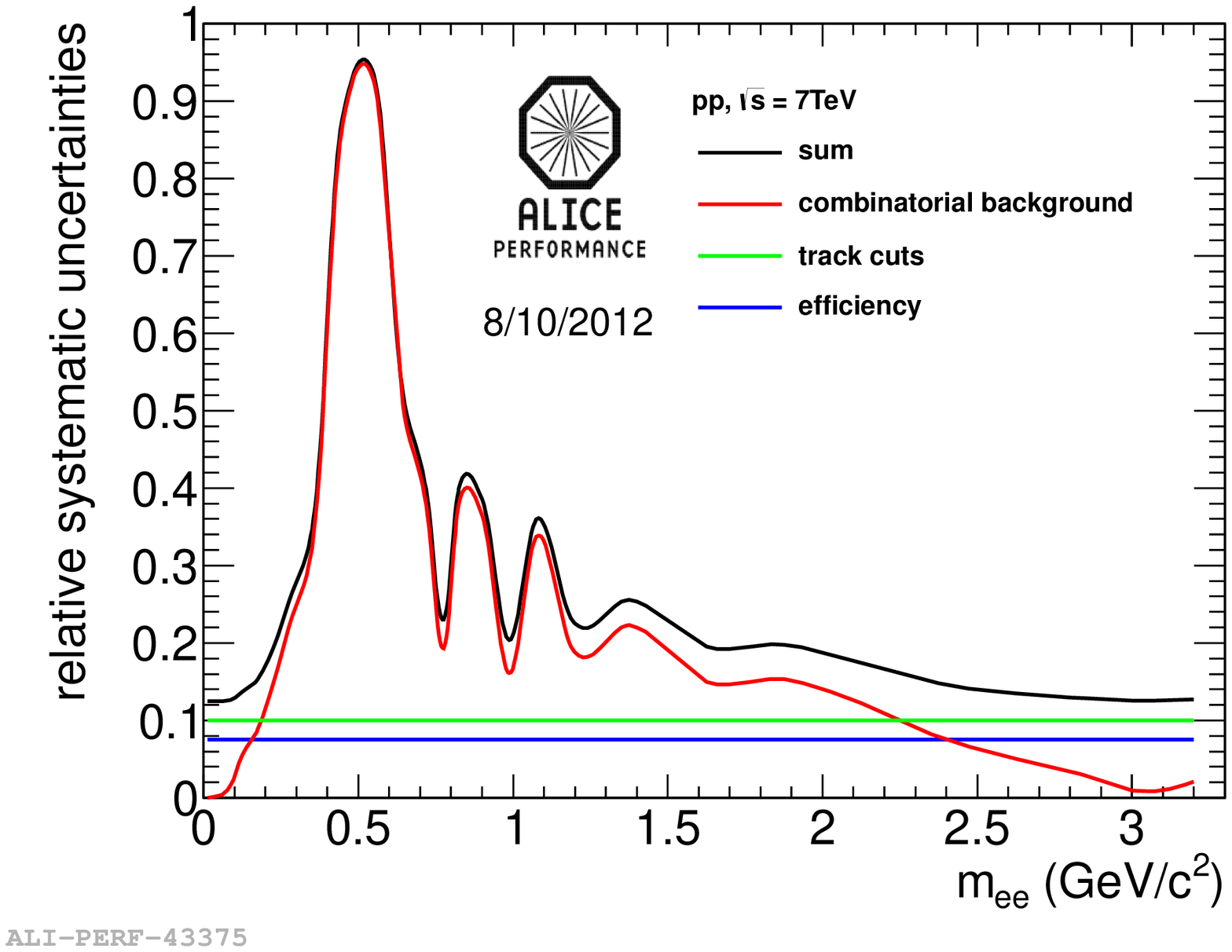}
 \caption{Sources of relative systematic uncertainties with their quadratic sum.}\label{fig:syst_uncertainty} 
\end{minipage} 
\end{figure}
The analysis for Pb--Pb collisions at $\sqrt{s_{NN}} = 2.76$~TeV is currently ongoing. Figure~\ref{fig:LeadLead} shows the signal to background ratio 
$S/B$ for central (0-10\%) Pb--Pb collisions in the pair transverse momentum range $1 < \pT^{\rm pair}<2$~GeV/$c$. The ratio reaches values of the 
order of $10^{-3}$, thus the systematic evaluation of the combinatorial background is the most important step of the analysis 
in the Pb--Pb collision system. 
\section{Results}
\begin{figure}[t]
 \centering
 \includegraphics[scale= 0.7]{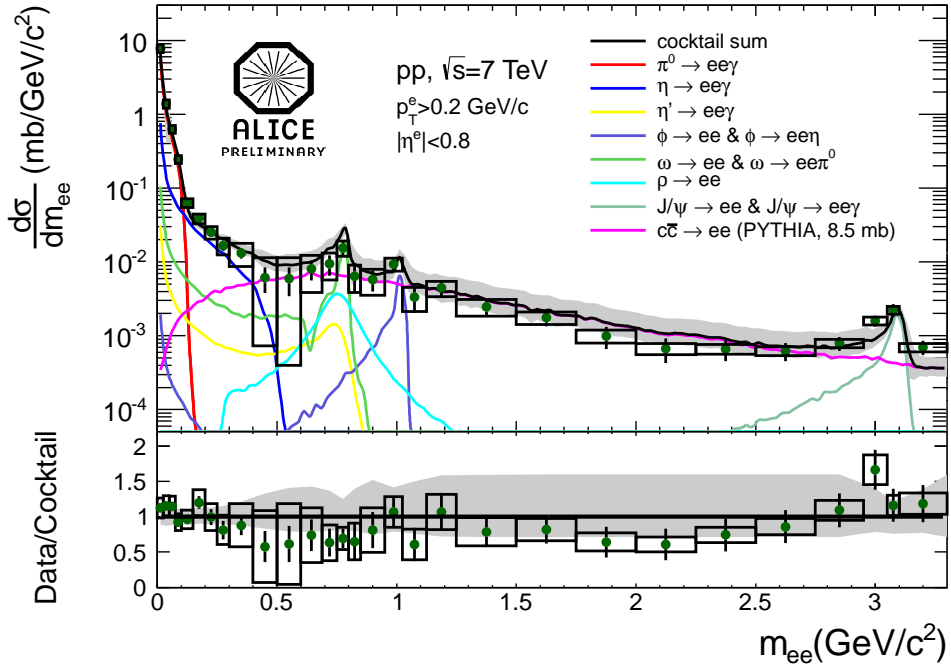}
\caption{Background subtracted invariant mass distribution in terms of total cross section compared to the hadronic cocktail for pp collisions 
at $\sqrt{s} = 7$~TeV. The boxes indicate the systematic uncertainty of the measurement, the grey band shown the systematic uncertainty of the cocktail.}\label{fig:cocktail_comparison} 
\end{figure}

The systematic uncertainty is dominated by the combinatorial background subtraction over a wide mass range of the dielectron continuum. This uncertainty 
depends on the $S/B$ ratio, 
shown in Figure~\ref{fig:pp_SoB}. The systematic uncertainties due to track cuts and the efficiency correction 
contribute significantly only at very low and high masses, where the $S/B$ ratio is large. 
The resulting relative systematic uncertainties are shown in Figure~\ref{fig:syst_uncertainty}. \\
Figure~\ref{fig:cocktail_comparison} shows the comparison of the dielectron continuum with 
the expected hadronic sources for pp collisions 
at $\sqrt{s} = 7$~TeV measured with the ALICE detector system. The hadronic cocktail agrees 
with the data within statistical and systematic uncertainties.\\
In the future, the proposed upgrade of the central barrel detectors~\cite{LoI}, especially of the ITS and TPC, will 
help to improve the low-momentum tracking efficiency
and also to detect electrons from secondary vertices from semi-leptonic heavy flavour hadron decays. This will be of special relevance for 
the measurement of Pb--Pb collisions.

\section*{References}
\bibliography{mkk_HQ2012_Proc}

\end{document}